Solar irradiance changes and photobiological effects at Earth's surface following astrophysical ionizing radiation events

Running title: Biological impact of ionizing events

Brian C. Thomas[1*], Patrick J. Neale[2], and Brock R. Snyder II[1]

[1]Washburn University, Department of Physics and Astronomy, Topeka, KS;

[*]Corresponding author: 1700 SW College Ave., Topeka, KS 66604; 785-670-2144; brian.thomas@washburn.edu

[2]Smithsonian Environmental Research Center, Edgewater, MD

Abstract:

Astrophysical ionizing radiation events have been recognized as a potential threat to life on Earth, primarily through depletion of stratospheric ozone and subsequent increase in surface-level solar ultraviolet radiation.  Simulations of the atmospheric effects of a variety of events (such as supernovae, gamma-ray bursts, and solar proton events) have been previously published, along with estimates of biological damage at Earth's surface.  In this work, we employed the TUV radiative transfer model to expand and improve calculations of surface-level irradiance and biological impacts following an ionizing radiation event.  We considered changes in surface-level UVB, UVA, and photosynthetically active radiation (visible light) for clear-sky conditions and fixed aerosol parameter values. We also considered a wide range of biological effects on organisms ranging from humans to phytoplankton. We found that past work overestimated UVB irradiance, but that relative estimates for increase in exposure to DNA damaging radiation are still similar to our improved calculations. We also found that the intensity of biologically damaging radiation varies widely with organism and specific impact considered; these results have implications for biosphere-level damage following astrophysical ionizing radiation events. When considering changes in surface-level visible light irradiance, we found that, contrary to previous assumptions, a decrease in irradiance is only present for a short time in very limited geographical areas; instead we found a net increase for most of the modeled time-space region. This result has implications for proposed climate changes associated with ionizing radiation events.

1. Introduction

Astrophysical ionizing radiation events have been recognized as a potential threat to life on Earth for decades. While there is the possibility for direct biological damage from redistributed radiation (Martin et al., 2009, Peñate et al., 2010), several studies have indicated that the greatest long term threat is from ozone depletion and subsequent heightened solar ultraviolet (UV) radiation (Reid and McAffee, 1978; Gehrels et al., 2003; Thomas et al., 2005; Thomas, 2009; Atri et al. 2014). Stratospheric ozone normally shields surface-dwelling life from harmful UV, particularly the UVB band, 280-315 nm. Atmospheric ionization, caused by high energy photons or cosmic ray particles, creates nitrogen oxides (most importantly NO and $NO_2$) which catalytically destroy ozone. Depending on the event's energy fluence and spectrum, this depletion can be severe and long lived, leading to greatly increased surface irradiation by solar UV (Thomas et al., 2005; Ejzak et al., 2007). It is known that organisms exposed to this irradiation experience harmful effects such as sunburn and even direct damage to DNA, proteins, or other cellular structures.

Simulations of the atmospheric effects of a variety of events have been completed, including gamma-ray bursts (both long-soft and short-hard types), supernovae (SNe), and solar proton events (SPEs). That work focused primarily on production of $NO_x$ (NO and $NO_2$) and subsequent depletion of stratospheric $O_3$ (Gehrels et al., 2003; Melott et al., 2005; Thomas et al., 2005, 2007, 2008, 2012, 2013; Ejzak et al., 2007). Estimates of biological damage following $O_3$ depletion have also been presented, using calculations of surface-level (solar) UVB combined with a DNA damage weighting function (Setlow, 1976; Smith et al., 1980).

Melott and Thomas (2011) reviewed a wide range of astrophysical events and concluded that supernovae, short-hard gamma-ray bursts (SHGRBs), and long-soft gamma-ray bursts (LSGRBs) (respectively) are the most likely astrophysical events to pose serious threats to life on Earth over a few 100 million year time scales. Determining these rates depends, in part, on estimates of the biological damage of varying ozone depletion levels. Melott and Thomas (2011) assumed that globally averaged depletion of 5% would have a noticeable biological impact, based on the fact that today's anthropogenic depletion is about this amount, while globally averaged depletion of 35% is likely to have mass-extinction level effects, based on results of previous modeling work.

A 35% global average decrease of $O_3$ in Earth's atmosphere is expected to have a major effect on terrestrial and marine life that is exposed to solar UV radiation. However, the calculations of UVB irradiance and use of relative DNA damage estimates in past work has inherent limitations; we would like to have both more accurate and more varied measures of the biological damage associated with such extreme $O_3$ depletion. Such improved results should make possible better estimates of the rates of threatening events.

The goal of this study, then, is to broaden and improve estimates of the biological impact on Earth's surface of astrophysical ionizing radiation events. To this end, we employed full atmospheric radiative transfer modeling, utilizing results of previously completed atmospheric modeling to produce surface-level spectral irradiance values in the UV and visible for various cases. These results were then used to compute biologically damaging exposure of a variety of impacts. We limited our modeling to clear-sky conditions (no clouds) and fixed aerosol parameter values.

While the atmospheric modeling utilized here represents realistic conditions, it does not include changes to clouds or aerosols that could potentially affect some chemistry that is of interest; we discuss this issue further in Section 2.1 below. The assumptions employed in radiative transfer modeling of clear-sky conditions and fixed aerosol parameters obviously limit our results to a simplified physical situation, which will not match all realistic possibilities. However, this choice greatly simplifies interpretation of our most important results and in particular allows us to focus on a "worst case" scenario.

2. Methods

2.1. Atmospheric modeling

Here, we use previously completed atmospheric modeling reported in Thomas et al. (2005). That paper presents results of modeling the effects of a long-soft gamma-ray burst with variation of many different parameters, including fluence, time of year of event, and location of event in latitude. Modeling reported in that work was performed by using the NASA Goddard Space Flight Center two-dimensional (latitude, altitude) time-dependent atmospheric chemistry and dynamics model (hereafter referred to as the "GSFC atmospheric model"). The model contains 65 chemical species and computes atmospheric constituents in a largely empirical background of solar radiation and galactic cosmic ray variations, with photodissociation included. The model includes heterogeneous processes (e.g., on polar stratospheric clouds), which are important for controlling ozone depletion and winds and small scale mixing. The model does not include dynamic feedback to the temperature fields and so cannot be used as a climate model. Extensive detail about the atmospheric modeling, including a full discussion of all included reactions and processes, as well as discussion of modeling uncertainties, can be found in the works of Thomas et al. (2005) and Ejzak et al. (2007).

A correlation between cosmic ray flux and clouds in Earth's atmosphere has been described (see, for instance, Marsh and Svensmark, 2000). However, this correlation has been called into question by Sloan and Wolfendale (2008) and Erlykin and Wolfendale (2011). Further, the claim of a correlation between climate change and spiral arm crossings (used to support the cosmic ray-cloud connection) by Shaviv (2002) has been effectively ruled out based on updated knowledge of Galactic dynamics (Overholt et al., 2009). The physical mechanism itself, involving enhanced formation of aerosols and nucleation sites, (Tinsley, 2000; Svensmark et al., 2009), while plausible, is not established (Wagner et al., 2001; Pierce and Adams, 2009). The CLOUD experiment at CERN has sought to establish such a mechanism, with inconclusive results (Duplissy et al., 2010, Kirkby et al., 2011).

While this possible effect on clouds and aerosols is relevant to our present study, with the large uncertainty surrounding the issue and no well-established physical mechanism, we could not accurately include such effects in atmospheric chemistry modeling. Furthermore, with no quantitative connection between ionization and aerosol/cloud characteristics, there is no way to realistically determine the appropriate parameter values for radiative transfer modeling. We therefore ignore any changes in aerosols or clouds proceeding from ionizing radiation in the atmosphere.

Ejzak et al. (2007) showed that $O_3$ depletion depends on total energy and spectral hardness, not duration of the event. For this study, changes in $NO_2$ and $O_3$ are the parameters of interest; therefore, the exact astrophysical event chosen is not important. Our results may be applied to any event with similar atmospheric consequences, including SNe and both types of GRBs. Further, our computational technique may be applied to any future modeling of events that cause significant atmospheric ionization leading to formation of $NO_2$ and depletion of $O_3$; the event parameters of relevance are energy fluence and spectral hardness.

We chose to focus our analysis here on the results of one particular case: an LSGRB with fluence 100 kJ m$^{-2}$, occurring over Earth's South Pole, in late June (roughly the Southern Hemisphere winter solstice). This choice is motivated by several factors. First, Melott and Thomas (2009) argued that a South-Polar burst fits well with what is currently known about the late Ordovician mass-extinction. Second, we wish to examine an extreme (but still realistic) event, with severe $O_3$ depletion; a polar burst isolates the ozone-depleting compounds to that hemisphere, which tends to result in greater localized $O_3$ depletion. Finally, the June case is a middle-ground example of maximum depletion for a polar burst. The maximum localized decrease in total vertical column density of $O_3$ for this case is about 70%, and the maximum globally averaged decrease in total vertical column density is about 32%, roughly the level of depletion assumed to be mass-extinction level in previous work (e.g., Melott and Thomas, 2011).

Figure 1 shows the percent difference in column density $O_3$ for runs with and without GRB ionization input, as a function of latitude and time after the event. Note that because the GRB was modeled as occurring over the South Pole, the $O_3$ depletion is concentrated in that hemisphere; we will focus further analysis to these Southern latitudes. The apparently anomalous "spikes" of increased $O_3$ in the extreme South-Polar region at roughly days 50-75 and 400-410 are caused by abrupt photolysis of $NO_2$ at the beginning of Polar spring, which generates large quantities of atomic O that briefly generates $O_3$. This feature is well understood, and more details can be found in the work of Thomas et al. (2005). Our analysis of biological effects will focus on mid-Southern-latitudes, and so we will not be concerned with this complication of $O_3$ production. Our focus on mid latitudes is motivated by previous work (e.g., Thomas et al., 2005) showing that, for a range of GRB cases over different latitudes and times of year, DNA-damage weighted irradiance is greatest at mid latitudes (about 20-50°).

Figure 2 shows the globally averaged percent difference in column density $O_3$ over 10 years following the event. Note that full recovery takes just over 10 years, and globally averaged depletion greater than 20% persists for just over 4 years.

2.2. Atmospheric Radiative Transfer Modeling

The primary goal of this study was to improve estimates of the biological impact of astrophysical ionizing radiation events. Results of the atmospheric modeling described above are passed to a radiative transfer model, producing surface-level spectral irradiance values between 280 nm and 700 nm (covering the UVB, UVA, and visible wavebands). These results are then used to compute exposure weighted for a variety of biological impacts, as described below. To simplify our conclusions and focus on "worst case" outcomes, we performed radiative transfer modeling assuming clear skies, with no clouds. We included realistic (but time-independent) aerosol parameter values, as described below.

Our primary tool for computing surface irradiance is version 5.0 of the publically available Tropospheric Ultraviolet and Visible (TUV) radiative transfer model, downloaded from http://cprm.acd.ucar.edu/Models/TUV/ (Madronich and Flocke 1997). Calculations are performed for atmospheric conditions following astrophysical ionizing radiation events, as described above. Several modifications have been made to TUV to directly use the profiles produced by the GSFC atmospheric model.

Previously published work (Thomas et al., 2005, 2007; Ejzak et al., 2007) included computed estimates of the increase in surface UVB under an ozone-depleted atmosphere using a simplified Beer-Lambert approach that included only absorption by a column of $O_3$ and neglected scattering and effects of other atmospheric constituents. We compared results from TUV to those using the Beer-Lambert approach, and found that the latter overestimates the UVB irradiance in every case considered by about a factor of 2 or less; the overestimate increases with solar zenith angle. These results are not surprising; since the Beer-Lambert method neglects scattering, we would expect to see higher surface level irradiance values.

It should be noted that results described by Thomas et al. (2005), Ejzak et al. (2007), and authors of related works did not directly report surface-level UVB values. Instead, relative DNA damage values were given (computed using the Setlow (1974) DNA damage weighting function). Therefore, past work did not consider absolute UVB or DNA damaging (weighted) exposure, but reported instead the normalized values to give a sense of the relative damage at any given location and time. This approach has the advantage of removing effects of uncertainties such as the simplified Beer-Lambert calculation. However, in the present work we wish to more quantitatively define the damaging exposure expected for marine phytoplankton and other biological impacts, and therefore will look at absolute irradiance values combined with more appropriate weighting functions. This requires the more accurate surface irradiance values produced by a full radiative transfer model.

2.2.1. Treatment of Aerosols and Selection of Aerosol Parameter Values

A potentially important variable affecting surface irradiance in the real atmosphere is the presence of aerosol particles. Our investigation focused on irradiation over the open ocean, at mid latitudes, and we therefore sought parameter values appropriate for this setting. As noted above, we restricted our modeling to clear-sky conditions, so we set the cloud optical depth to zero in TUV.

TUV reads in a total aerosol optical depth (AOD) value, along with values for single scattering albedo (SSA) and Angstrom coefficient ($\alpha$). In addition, a wavelength-independent asymmetry factor ($g$) is set by default to be 0.61. The default aerosol profile in TUV is for continental conditions. We modified the default TUV aerosol profile to match that used for clean marine conditions in another widely used radiative transfer model, the System for Transfer of Atmospheric Radiation (STAR) (Ruggaber et al., 1994). Figure 3 shows the aerosol profile used.

We searched the literature to determine appropriate ranges for the aerosol parameters in TUV, for clean air over the open ocean at mid latitude ("clean marine" conditions). Default values in the model are for continental conditions, with AOD = 0.235 (at a wavelength of 550 nm), SSA = 0.99, and $\alpha$ = 1.0. Most measurements of optical depth are given at a wavelength of 500 nm. Kedia and Ramachandran (2009) listed a range of values for a "maritime clean" model from 0.04 to 0.13. They also quoted measurements from the AERONET program, which range from 0.07 to 0.14. We examined data from AERONET for sites at Bermuda, Midway, Coconut Island, Tahiti, Guam, Azores, Prospect Hill, and Ascension Island. Level 2.0 data was retrieved from the AERONET website (http://aeronet.gsfc.nasa.gov/). For each location, we recorded the annual average for the year 2001 of the 500 nm AOD and 870-440 nm Angstrom parameter values. The range in AOD values for the sites listed is from 0.063 to 0.187. Mulcahy et al. (2009) gave an average "clean marine" AOD value of 0.14 ± 0.06.

Taking an average of the extreme range represented by the various ranges above (from 0.04 to 0.20) gives 0.12, at 500 nm. TUV requires input of the optical depth at 550 nm. The optical depth at a given wavelength is given by $\tau_1 = \tau_2 (\lambda_2 / \lambda_1)^\alpha$, where $\tau_i$ is the optical depth at wavelength $\lambda_i$ and $\alpha$ is the Angstrom coefficient. As discussed below, we take $\alpha = 0.4$, which gives AOD = 0.116 at 550 nm.

Petters et al. (2003) and Kazantzidis et al. (2001) gave single scattering albedo values of about 0.89. Hatzianastassiou et al. (2004) gave a value of 0.96. Mulcahy et al. (2009) described clean marine conditions as being dominated by "relatively large supermicron particles, such as sea salt," which are highly reflective. We therefore chose to leave SSA at its default value of 0.99.

For the Angstrom coefficient, Kedia and Ramachandran (2009) listed a range of values for a "maritime clean" model from 0.2 to 0.93, for the wavelength range 440 to 870 nm. Data from AERONET for the sites listed above has a range from 0.290 to 0.912. Mulcahy et al. (2009) gave an average "clean marine" value of 0.40 +- 0.29. We adopted this average value, setting a = 0.4 in TUV.

The asymmetry factor in TUV is by default set to 0.61 at all altitudes and wavelengths. We have modified this to 0.75 (retaining altitude and wavelength independence), based on values given by Hatzianastassiou et al. (2004) and Papadimas et al. (2009) for clean marine conditions.

Overall, then, we take the following values as best for clean marine conditions in TUV: AOD = 0.116 (at 550nm), SSA = 0.99, $\alpha$ = 0.4, g = 0.75.

We performed analyses to determine how sensitive modeling results are to changes in aerosol parameters. While we do not report those details here, we note some general conclusions. First, for any result of interest to us, a physically realistic variation in the Angstrom coefficient has no significant effect. Single scattering albedo shows a larger difference, but since we wish to restrict our attention to clean marine conditions (without, for instance, large amounts of black carbon), we believe that a fixed value of 0.99 is realistic. For total aerosol optical depth, we see small variation in total irradiance for all wavebands (UVB, UVA, and PAR) across the AOD range of interest. Since our primary interest is the change in surface-level total irradiance and its biological impact, potential variability in AOD is not of great concern. Our choice of an average value should suffice for general purposes and in particular for comparisons of the impact of differing ozone column densities and other parameters, which is our primary purpose. Finally, we note that modifying the default, continental aerosol profile in TUV to a clean marine profile resulted in less than 1% difference in spectral difference in spectral irradiance in the UV and visible.

We also checked the sensitivity of surface irradiance to surface albedo value. An application of this work (evaluating water-column photosynthesis productivity of phytoplankton, reported elsewhere) requires computations involving penetration of light into ocean water. Ocean surface albedo varies depending on several factors including solar zenith angle, total aerosol optical depth, wind speed, and ocean chlorophyll concentration (Jin et al. 2004). Of these parameters, solar zenith angle has the largest effect, with the albedo value ranging from about 0.03 to about 0.3, depending on specific values of the parameters considered. In TUV, albedo is set to a single constant value by default. Leaving all other parameters fixed, we compared results for albedo values of 0.02 and 0.44 (the extreme range reported in Jin et al. 2004). We found a 22% difference in total UVB irradiance. A 0.1 difference in albedo gives a 5% difference in total UVB.

To improve the accuracy of our results, we modified TUV to incorporate a dynamic surface albedo, utilizing the look-up table described by Jin et al. (2004). Data and a read-in subroutine were obtained from http://snowdog.larc.nasa.gov/jin/getocnlut.html. The read-in subroutine was used as a template to modify the setalb subroutine in TUV. This modified subroutine takes values from the main program for solar zenith angle, as well as total aerosol optical depth. In addition, values are specified (manually) for wind speed and ocean chlorophyll concentration; these values are not normally a part of the TUV model. While there is of course variability in these values in nature, we have chosen representative values of 7.8 m s$^{-1}$ for wind speed (an annual global average of values from SCOW, obtained from http://cioss.coas.oregonstate.edu/scow/) and of 0.20 mg m$^{-3}$ for chlorophyll concentration (a global average value, given in Jin et al. 2004).

As noted in Section 1, out assumptions of clear-sky conditions and fixed aerosol parameters do limit our results to a simplified physical situation. This allows us to focus on our main results and present a "worst case" situation for an ionizing radiation event. In addition, with no realistic way of knowing how aerosol parameters may be affected by ionizing radiation in Earth's atmosphere (see Section 2.1), a fixed set of values seems most appropriate for this study; as noted above, realistic variations in the TUV aerosol parameters do not significantly affect our primary results.

Finally, it should be noted that our main results (Section 3) are given as relative values, comparing results in the GRB case versus results for a normal atmosphere case. Both cases are run with exactly the same TUV settings, except for $NO_2$ and $O_3$ number density profiles. Therefore, variations in quantities such as the aerosol parameters should not be a significant factor in those relative values.

2.2.2. Adapting TUV for use with Atmospheric Modeling Results

Some modifications were made to TUV for easier use with results from the GSFC atmospheric model. First, TUV was modified slightly to read in an altitude profile of $O_3$ number density from the ground to 90 km (with 46 levels). This profile is an output of atmospheric modeling of a particular event when using the GSFC model, and includes both altitude values and corresponding $O_3$ number density values. We have similarly modified TUV to read in an altitude profile of $NO_2$. This constituent has a strong absorption band centered around 400 nm and is present in greatly enhanced quantities following an ionizing radiation event. We are interested in considering the impact on primary productivity, and so reductions in photosynthetically available radiation are important; in addition, change in UVA can be significant for photosynthesis inhibition. When $O_3$ and $NO_2$ profiles are read in, the total column density is computed from the profile, with the option to instead scale to an input value. We have further modified TUV to output both direct and diffuse spectral irradiance across the range from 280 to 700 nm, with 1 nm resolution.

With changes in $O_3$ concentration, changes in the temperature profile of the atmosphere occur. These changes could potentially impact surface UV irradiance, since $O_3$ absorption is temperature dependent. The GSFC model does not include dynamic feedback to temperatures, and so we cannot include these changes in the radiative transfer modeling. However, previous studies (Thomas et al. 2005) have indicated that temperature changes are likely to be small (around 10 K). We have conducted tests with TUV with modified temperature profiles. The $O_3$ profiles associated with an ionizing radiation event show decreased concentrations in the stratosphere, but increased concentrations in the upper troposphere. We therefore modified the standard temperature profile with an increase of 10 K between 10 and 15 km altitude, and a decrease of 10 K between 20 and 30 km altitude. Irradiance is changed by less than 1% in this case, over all wavelengths of interest; a similar change of 50 K yields approximately 1% difference. We therefore conclude that lack of temperature feedback is not a significant source of error here.

2.3. Modeling of photobiological impacts

Most previous studies seeking to understand the impact of astrophysical ionizing radiation events on life on Earth have relied on simple measures such as the increase in UVB alone, or DNA damage due to enhanced UVB (Thomas et al., 2005; Ejzak et al., 2007; Melott and Thomas, 2009, 2011; Martin et al. 2009), though Peñate et al. (2010) also considered the impact of the short-term UV "flash" from a GRB on oceanic photosynthesis. Computing changes in UVB or DNA damaging exposure provides a simple way to evaluate the potential severity of biological impacts of various events. It is also useful in exploring features such as the spatial extent and duration of the most intense impacts. However, changes in UVB irradiance or DNA damaging exposure alone is overly simplistic when trying to evaluate just how damaging a particular event may be to Earth's biosphere as a whole. In the first place, UVB is not the only biologically active waveband affected by atmospheric changes that follow an ionizing radiation event; the UVA and PAR bands are also affected, as discussed above. Further, these different forms of radiation can have complicated, and interacting, effects, and those effects may be very different for different organisms. In particular, DNA damaging exposure estimates overlook the fact that most organisms have protection and repair mechanisms, some of which are affected by UVA and PAR irradiance (Williamson et al. 2001; Bancroft et al. 2007).

An important goal of this study was to use our improved surface-level irradiance calculations to look much more broadly and accurately at the biological impacts we may expect following an intense astrophysical ionizing radiation event. While we focus here on one particular kind of radiation event, our results will be applicable to a wide range of events (as discussed in Sections 1 and 2.1), and our computational approach can be applied to any event of interest.

Many different photobiological impacts of UV radiation have been identified. The most obvious for humans are sunburn (erythema) and skin cancer, which have also been identified in other organisms, such as whales (Martinez-Levasseur et al., 2011) and fish (Sweet et al., 2012). Other work has focused on aquatic life of various types (e.g. Boucher and Prezelin 1996; Hader, 1997; Kouwenberg et al., 1999; Hader et al., 2007; Bancroft et al., 2007), amphibians (Blaustein et al., 1994; Blaustein and Kiesecker, 2002), terrestrial plants (e.g. Flint and Caldwell, 2003a,b; Wu et al. 2009; Kakani et al., 2003; Searles et al., 2001), and terrestrial ecosystems as a whole (Ballare et al., 2011; Caldwell et al., 1998, Bjorn et al., 1996).

The TUV model includes a variety of built-in biological weighting functions (BWFs) that may be used to compute effective exposure for impacts such as direct DNA damage, sunburn and skin cancer in humans, cataracts in pig eyes (a good model for human eyes), plant damage, and phytoplankton photosynthesis inhibition. In Section 3.2, we present results of applying several BWFs to our surface irradiance results.

3. Results

3.1. Surface-level irradiance

In Figure 4, we present ratios of irradiance in the GRB case versus that in a normal background case for the first two years following the event. Note that these figures are restricted to a latitude range from 5°N to 65°S; the motivation for this range of focus is discussed in Section 2.1. The ratio is computed for noon time irradiance at each location, under clear skies (no clouds). Figure 4 shows the ratios of UVB (280-315 nm), UVA (315-400 nm), and photosynthetically active radiation (PAR, 400-700 nm). In addition, Figure 5 shows spectral ratios from 295 nm to 700 nm, at 35°S for two specific days after the event. Several features may be noted in these results. First, UVB is everywhere higher in the GRB case (seen in Figure 4 and the top panels of Figure 5), as may be expected due to the depletion of $O_3$. (Note in Figure 5 that the top panels show the log ratio of irradiance in the UVB part of the spectrum.) Second, UVA is everywhere lower in the GRB case (seen in Figure 4 and the bottom panels of Figure 5); this is due to increased $NO_2$ column density, which absorbs strongly between about 320 nm and 480 nm. Finally, in the GRB case PAR is initially lower and then becomes higher (seen in Figure 4 and the bottom panels of Figure 5).

This change in PAR is due to a competition between increased $NO_2$ column density and decreased $O_3$ column density. $NO_2$ absorbs in the short wavelength (approximately 400-500 nm) end of the PAR region, while $O_3$ has an absorption band centered around 600 nm and extending across most of the PAR region. These competing effects can be seen most clearly in the bottom panels of Figure 5. Initially, $NO_2$ is greatly enhanced by the induced atmospheric ionization and then slowly is reduced by conversion into other species such as $NO_3$; the greatest $O_3$ depletion is somewhat delayed (see Figure 6), due primarily to seasonal effects (see Section 2.1). Therefore, initially the absorption of PAR by $NO_2$ dominates, leading to a decrease in the total irradiance; later, decreased absorption by $O_3$ leads to a net increase in PAR. It should also be noted that, while broad-band PAR is initially decreased by $NO_2$, there is still a small increase across part of the PAR wavelength range. This is due to the reduced absorption by $O_3$ that begins immediately following the event (which then worsens over the next few years before recovering).

These results have implications for different biological impacts. UVB is uniformly damaging for organisms. UVA can be damaging, but also plays a role in photorepair of damage caused by UVB (Williamson et al. 2001; Bancroft et al. 2007). PAR is of course important in photosynthesis, but also can contribute to photosynthesis inhibition. Therefore, the overall result is complicated – UVB will cause damage, but increased PAR may help offset that. On the other hand, UVA is important for photorepair mechanisms, and so reduced UVA may lead to enhanced damage caused by UVB. We do not attempt to examine all of these details here. Instead, we use BWFs that incorporate all relevant wavelengths and then consider overall biological results for each BWF.

Other work on the terrestrial effects of a GRB has considered the possibility of a reduction in PAR due to $NO_2$ absorption, and speculated that this change could cause a cooling in the climate (Melott et al., 2005; Thomas et al., 2005; Martin et al., 2010). As we have seen here, there is indeed a reduction in PAR initially. Reductions up to 75% are seen in our results, but only at the most extreme southern latitudes, lasting just a few months, varying seasonally. Figure 7 shows the ratio of PAR in the GRB case versus the normal case at latitudes 75°S, 65°S, 55°S, and 45°S. Combined with Figure 4, we can see that wide-spread reductions in PAR are limited to a few percent over the first 250 days or so, with an increase in PAR at most latitudes following that time, and an increase (or no change) at all latitudes after about the first year. These results to do not allow us to draw firm conclusions about climate change, since such changes must be evaluated in the context of all atmospheric and irradiance changes associated with an astrophysical ionizing radiation event. However, these results appear to make a global (or even regional) cooling episode much less likely than was previously assumed.

3.2. Surface-level biological impacts

Our primary interest in this work is the impact of changes in surface-level irradiance on the biosphere. To that end, we employed several built-in BWFs included in the TUV model. These weighting functions quantify the effectiveness of a range of wavelengths on certain biological processes, such as DNA damage, photosynthesis, plant growth, or skin damage (erythema). All computations are done for noon time irradiance at each location, under clear skies (no clouds).

The Setlow (1976) relative DNA damage weighting function is a commonly used BWF; it was used in previous work discussed above (e.g., Thomas et al., 2005). This BWF is included in TUV. The top panel of Figure 8 shows the ratio of irradiance weighted for DNA damage in the GRB case versus that in a normal background case, for four years following the event, again restricted to a latitude range from 5°N to 65°S. Note that there is more than a 5-fold increase compared to a normal background level, across most of the latitude range of interest. The value varies seasonally, as may also be seen in previous figures of, for instance, UVB increase. This seasonal variation is due to seasonal changes in both $O_3$ column density (see Figure 1) and solar zenith angle (with a smaller angle resulting in greater irradiance).

Previous work in which this BWF was used presented results in terms of values normalized by the annual global average under normal atmospheric conditions; we present a similar result for this work in the middle panel of Figure 8. In previous work (Thomas et al., 2005), the computed DNA damaging irradiance at any given point was a maximum of about 15 times the global annual average; in this work that maximum is about 11. A smaller value is not unexpected since surface level UVB irradiance is somewhat smaller in this study (due to the more complete radiative transfer modeling). Moreover, even this value is a conservative maximum since it was calculated at noon for clear sky conditions. Still, even with differences in computation technique and resulting differences in surface level irradiance, the ratio of DNA damaging irradiance relative to the global annual average in a normal atmosphere case is quite similar to previous results.  This indicates a certain robustness in our previous results, despite the simplified computational technique.

As in previous work, we again find that the greatest increase in DNA damaging irradiance is located at mid-latitudes.  However, one may note that the location in latitude of the maximum increase does depend on which ratio we choose; the top and middle panels of Figure 8 show the maximum values at somewhat different latitudes, with the point-to-point ratio of the GRB case versus the normal case showing a maximum at somewhat higher (Southern) latitudes than for the ratio of the GRB case to the global annual average.  As may be expected, we see recovery occurring slowly over a few years, tracking the return of stratospheric $O_3$ to normal values (see Figure 1).

One quantity of interest to human populations when discussing ground-level UV is the "UV Index" (WMO 1994).  This index is intended to be a simple way to communicate with the public the danger posed by UV radiation at any given time and place.  It is based primarily on a weighting function for damage to human skin (erythema).  The bottom panel of Figure 8 shows the UV Index as computed by the TUV model for the GRB case.  UV Index values of 11 and above indicate "extreme risk of harm from unprotected sun exposure," and people are encouraged to take all precautions against sun exposure.  As can be seen in the bottom panel of Figure 8, the UV Index values in our modeling reach up to 29 and are indeed significantly greater than 11 for most of the region of interest. This indicates a severe threat to human populations, in the form of sunburn and increased risk of skin cancer, lasting several years (varying seasonally).

To quantify the risk to humans in more detail, we present in Figure 9 results for irradiance weighted with BWFs for human erythema (Anders et al., 1995), skin cancer (de Gruijl and van der Leun, 1994), non-melanoma skin cancer (CIE, 2006), and cataracts (using a pig model; Oriowo et al., 2001). This figure shows the ratio of values for the GRB case to values for the normal atmosphere case. Note first that each panel of this plot has a different maximum value; this allows comparison of the time-space features of the increase in damage. Figure 10 shows ratio values at 45°S latitude, where most of the values shown are at a maximum; this plot allows easier comparison of the range of ratio values for different BWFs. Note that there is a rather wide range of ratio values, with the de Gruijl and van der Leun (1994) skin cancer BWF yielding, at maximum, nearly an order of magnitude greater increase in damaging exposure than that for non-melanoma skin cancer and for cataracts.

One goal of this work was to evaluate the impact of ionizing radiation events on primary productivity. This was motivated in part by an effort to understand how such events may have contributed to mass extinction events that are evident in the fossil record (Melott et al., 2004; Thomas et al., 2005; Melott and Thomas, 2009). We presume that impacts on the base of the food chain would have major consequences for overall biodiversity, potentially triggering, or at least contributing to, extinction events. The TUV model includes several BWFs for primary producers, including inhibition of carbon fixation in Antarctic phytoplankton (Boucher et al., 1994), inhibition of photosynthesis in the diatom *Phaeodactylum* sp. and the dinoflagellate *Prorocentrum micans* (Cullen et al., 1992), and inhibition of growth of higher plants (Flint and Caldwell, 2003a,b), specifically oat (*Avena sativa* L. cv. Otana) seedlings. In Figure 11, we present results of applying these BWFs, again as ratios of the computed result in the GRB case to that under normal conditions; here again each panel has a different maximum value allowing comparison of time-space features. Figure 12 shows ratio values at 45°S latitude, where most of the values shown are at a maximum, again allowing for easier comparison of the range of ratio values for different BWFs. Here the range of ratio values is smaller than for the human-impact results, with about a factor of 4 separating the maximum values for the most different BWFs.

Some general features of these BWF results may be noted. We again find that the greatest increase in damaging irradiance is located at mid-latitudes, with recovery occurring slowly over a few years, tracking the return of stratospheric $O_3$ to normal values. In the human-impact results, the greatest increase in damaging irradiance is seen for the de Gruijl and van der Leun (1994) skin cancer BWF with values ranging between approximately 3 and 12 times the normal background values for the entire time period shown (about 4 years). Interestingly, irradiance weighted non-melanoma skin cancer shows only a modest increase, less than a factor of 2 at maximum.

In the primary productivity results there is a smaller difference between the BWFs. The greatest increase in weighted irradiance occurs with the Boucher et al. (1994) BWF for inhibition of carbon fixation in Antarctic phytoplankton, with values ranging between about 2 and about 5 times the normal background values. The smallest increase in weighted irradiance occurs for the two inhibition of photosynthesis BWFs (Cullen et al., 1992), with greater increase for the diatom *Phaeodactylum* sp., and a rather small (less than a factor of 2) increase for the dinoflagellate *Prorocentrum micans*. Interestingly, the second greatest increase in weighted irradiance occurs for the Flint and Caldwell (2003a,b) BWF for inhibition of growth in oat seedlings. This result is significant from the point of view of human populations, since it indicates that a significant impact on food crops is likely, lasting several seasons.

4. Discussion and Conclusions

We have presented results of improved and expanded modeling of the surface-level irradiance and subsequent biological impact of a GRB, which may also be applied to other astrophysical ionizing radiation events. We find that previous work overestimated the surface-level UVB irradiance by about a factor of 2; however, values for the relative change in DNA damaging irradiance here are not significantly smaller than those previously reported. This is an important result that indicates previous studies are reliable, at least roughly, in evaluating the potential biological impact of ionizing radiation events. The simpler approach used in previous work is faster and easier to implement, so we conclude that it is still useful, at least in giving a general sense of the magnitude of potential threats to life on Earth (using relative DNA damage as a proxy).

Our surface-level irradiance results also have implications for suggestions in past work (Melott et al., 2005; Thomas et al., 2005; Martin et al., 2010) that the climate could cool following the production of $NO_2$ and accompanying reduction in visible light. That work neglected the compensating effect of reduced $O_3$, which actually leads to a net increase in visible light (PAR) over the long term. Net decreases in PAR are limited in area and duration, and hence it appears that climate cooling suggested in earlier work may be unlikely. We make no attempt here to further examine potential climate changes, which are complicated and affected by many factors related to atmospheric and irradiance changes. Any conclusions about associated climate changes will require sophisticated modeling that includes all atmospheric and irradiance changes following an ionizing radiation event, which is well beyond the scope of this work.

Besides improving the modeling of surface-level irradiance, we also sought to expand estimates of the biological consequences beyond simple DNA damage estimates. The biologically weighted irradiance results presented in Section 3.2 show that the impact of $O_3$ depletion following astrophysical ionizing radiation events can be highly variable, even for similar biological measures (such as skin cancer or productivity of marine phytoplankton), and significant increase in weighted irradiance for one organism or for one particular type of biological effect does not necessarily indicate a similar increase in weighted irradiance for all biological effects. This complicates the question of whether a particular level of $O_3$ depletion will lead to widespread catastrophe for Earth's biosphere. That said, we may still conclude that biological damage would be widespread and significant for many or most organisms. Extinction events always display a range of impacts on different groups of organisms, with some being completely wiped out, others barely surviving, and still others emerging relatively unscathed (Jablonski, 2005); our results exhibit this variability.

Importantly, our results also show that significant effects can occur at all levels of organism-complexity; humans, higher plants, and phytoplankton all show a significant increase in damage following $O_3$ depletion. Previous work has generally assumed that major $O_3$ depletion would hit primary producers especially hard, potentially leading to a collapse of the food web. These results support the idea that primary producers would be significantly impacted, but that impact is not uniform between various species. Additionally, we see significant impact on higher-complexity organisms, indicating that a food-web collapse is not necessary for there to be potential for extinction at higher trophic levels.

The variability of biological results motivates future use of yet more BWFs for more organisms and other impacts. In addition, these results indicate that future work should consider the broader ecological impact of irradiance changes. An ecosystem is a complex network of interactions, and determining whether a mass extinction would indeed result from a particular level of $O_3$ depletion will require examining biological damage in this context, as well as considering the implications of potential climate change or other terrestrial changes following the event.

We are aware of two studies that sought to evaluate the potential ecosystem impact of major ozone depletion events. Martin et al. (2010) considered phytoplankton mortality following an increase of surface-level UV irradiance and used the Comprehensive Aquatic Simulation Model (Amemiya et al., 2007) to study the impact on a lake ecosystem. This study is limited, however, by uncertainties and relative lack of data on the mortality caused by increased UV. In addition, changes in PAR are not considered. A complication recognized by Martin et al. (2010) is that the recovery of the atmosphere following an ionization event involves deposition of nitrate. While there is some possibility for stress caused by acidification, Melott et al. (2005) and Thomas and Honeyman (2008) showed that the most likely result is actually an increase in primary productivity due to the nitrate acting as a "fertilizer." This complication has not been included in any attempts to model ecosystem impacts.

Glassmeier et al. (2009) considered the effect of changes in surface-level UVB for the case of increased cosmic ray flux (which also causes depletion of stratospheric $O_3$). They also used a model of aquatic phytoplankton population. Here, too, changes in PAR are not considered; UVA changes are also omitted, and the impact of UVB is based solely on whether the irradiance passes a threshold for inhibition of photosynthesis.

While both of these studies are limited, they represent a starting point for evaluating the broader ecological impact of ionizing radiation events. A conclusion of both studies is that attenuation of light in the water column, especially due to the presence of dissolved organic matter, plays an important role in the overall impact on a phytoplankton community. Results presented here are all surface-level; future work will consider the transmission of light into the water column and the integrated effect on primary productivity in the oceans for some important phytoplankton species, providing further information on how the base of the food web may be impacted.

Overall, our results confirm that major biological impacts would be associated with severe $O_3$ depletion following an astrophysical ionizing radiation event. Importantly, we find that such impacts are variable across different organisms and effects considered. In addition, we find that the entire UVB, UVA, and PAR regions of the spectrum should be included in computing biological effects. Future work in this area should include more measures of biological impacts, with the ultimate goal of understanding the broader ecological impact.

Acknowledgements

The authors thank Sasha Madronich for assistance with using and modifying the TUV code. This work has been supported by the National Aeronautics and Space Administration under grant No.s NNX09AM85G and NNX14AK22G, through the Astrobiology: Exobiology and Evolutionary Biology Program. Computational time for this work was provided by the High Performance Computing Environment (HiPACE) at Washburn University; thanks to Steve Black for assistance with computing resources. The authors thank two anonymous reviewers for helpful comments that improved the manuscript.

Author Disclosure Statement: No competing financial interests exist.

Häder, D.P., Kumar, H.D., Smith, R.C., and Worrest, R.C. (2007) Effects of solar UV radiation on aquatic ecosystems and interactions with climate change. *Photochem. Photobio. Sci.* 6:267–285. DOI: 10.1039/b700020k

Hatzianastassiou, N., Katsoulis, B., and Vardavas, I. (2004) Global distribution of aerosol direct radiative forcing in the ultraviolet and visible arising under clear skies. *Tellus B* 56:51-71.

Jablonski, D. (2005) Mass extinctions and macroevolution. *Paleobiology* 31:192-210.

Jin, Z., Charlock, T.P., Smith, W.L., Jr., and Rutledge, K. (2004) A parameterization of ocean surface albedo. *Geophys. Res. Lett.* 31:L22301, doi:10.1029/2004GL021180.

Kakani, V.G., Reddy, K.R., Zhao, D. and Mohammed, A.R. (2003) Effects of Ultraviolet-B Radiation on Cotton (*Gossypium hirsutum* L.) Morphology and Anatomy. *Annals of Botany* 91:817-826.

Kazantzidis, A., Balis, D.S., Bais, A.F., Kazadis, S., Galani, E., Kosmidis, E. (2001) Comparison of Model Calculations with Spectral UV Measurements

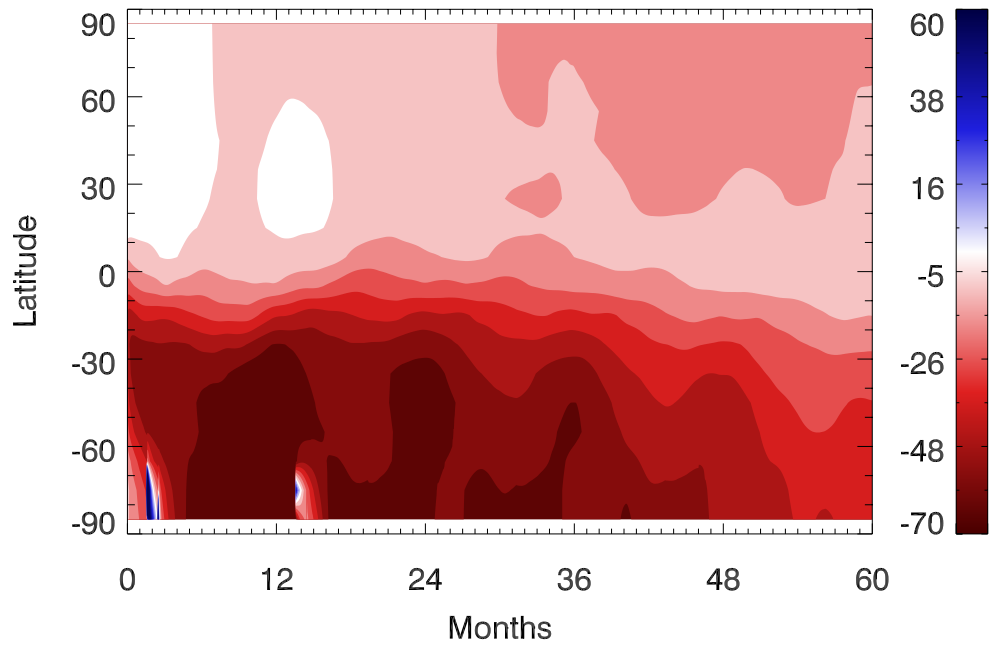

Figure 1

Pointwise percent difference in column density of $O_3$ (comparing runs with and without GRB). The GRB occurred at time 0, over the South Pole.

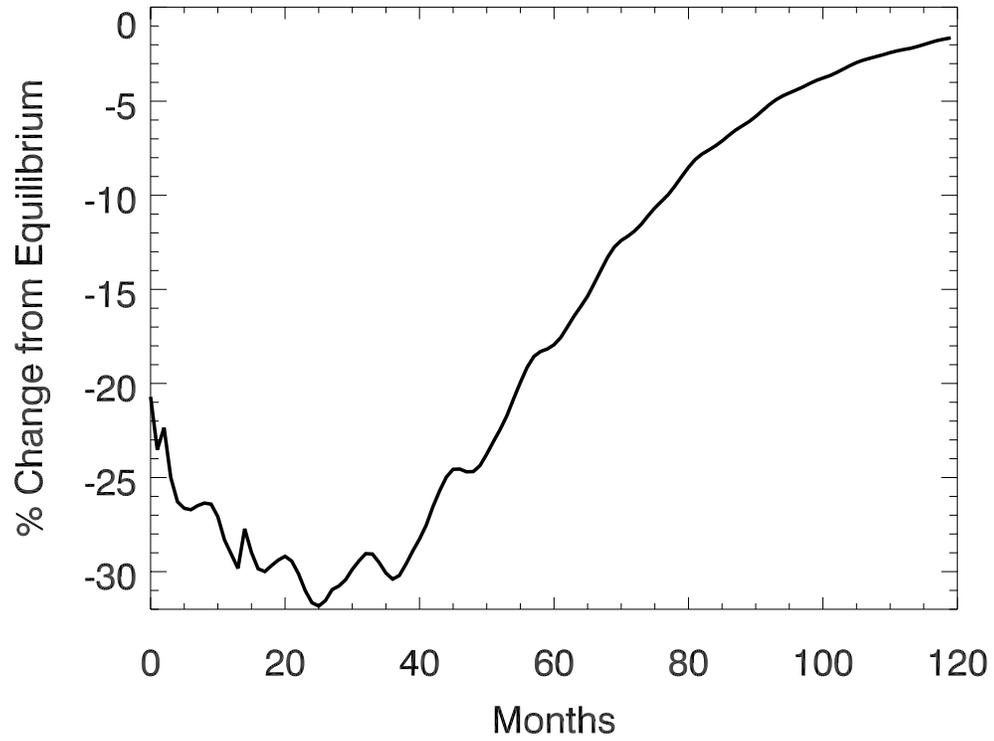

Figure 2

Percent change (comparing runs with and without GRB) in globally averaged column density of $O_3$.

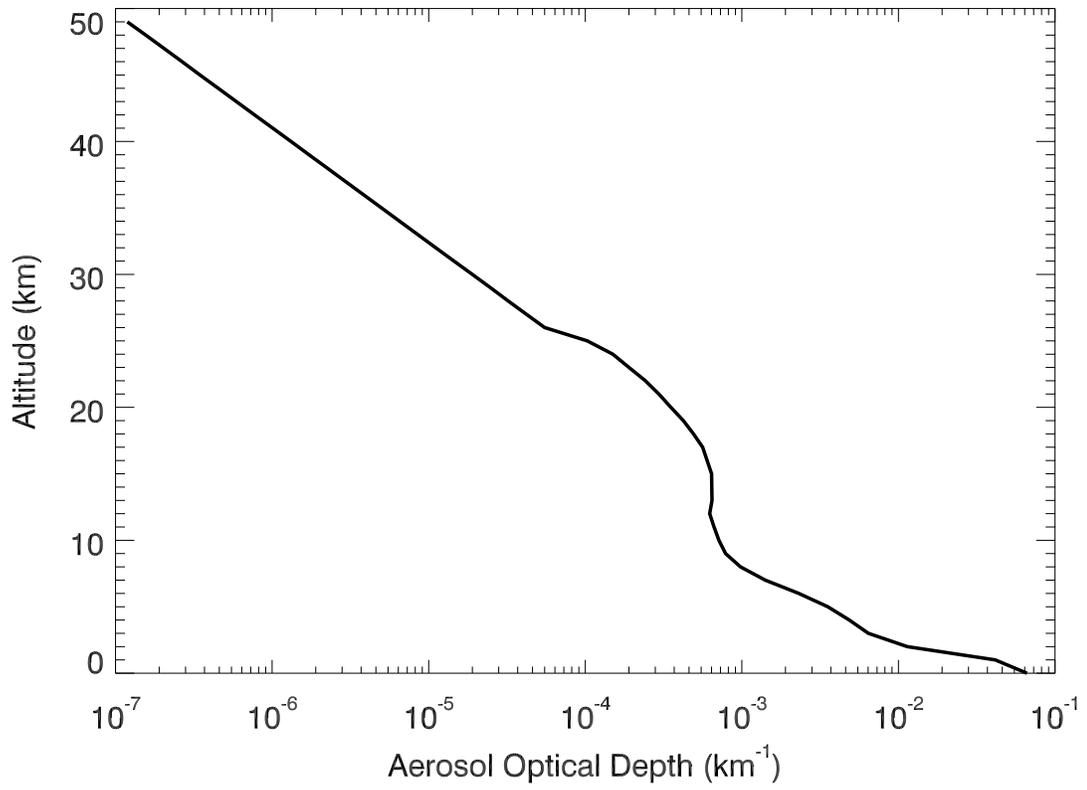

Figure 3

Aerosol profile used in TUV radiative transfer modeling for this study.

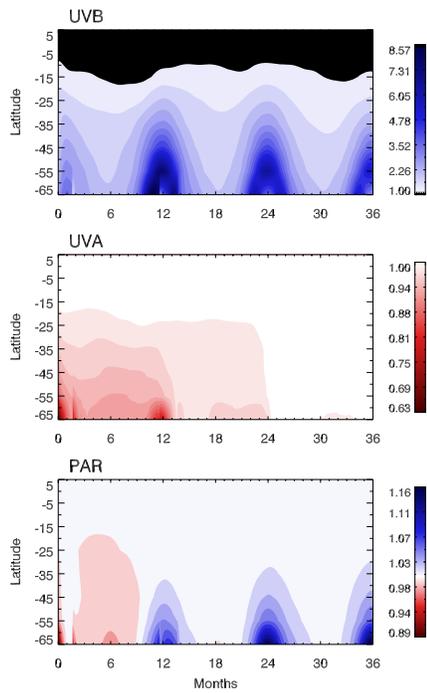

Figure 4

Pointwise ratio of irradiance at Earth's surface in GRB case versus normal case (where the GRB occurred at time 0), for UVB (top panel), UVA (middle panel), and PAR (bottom panel).

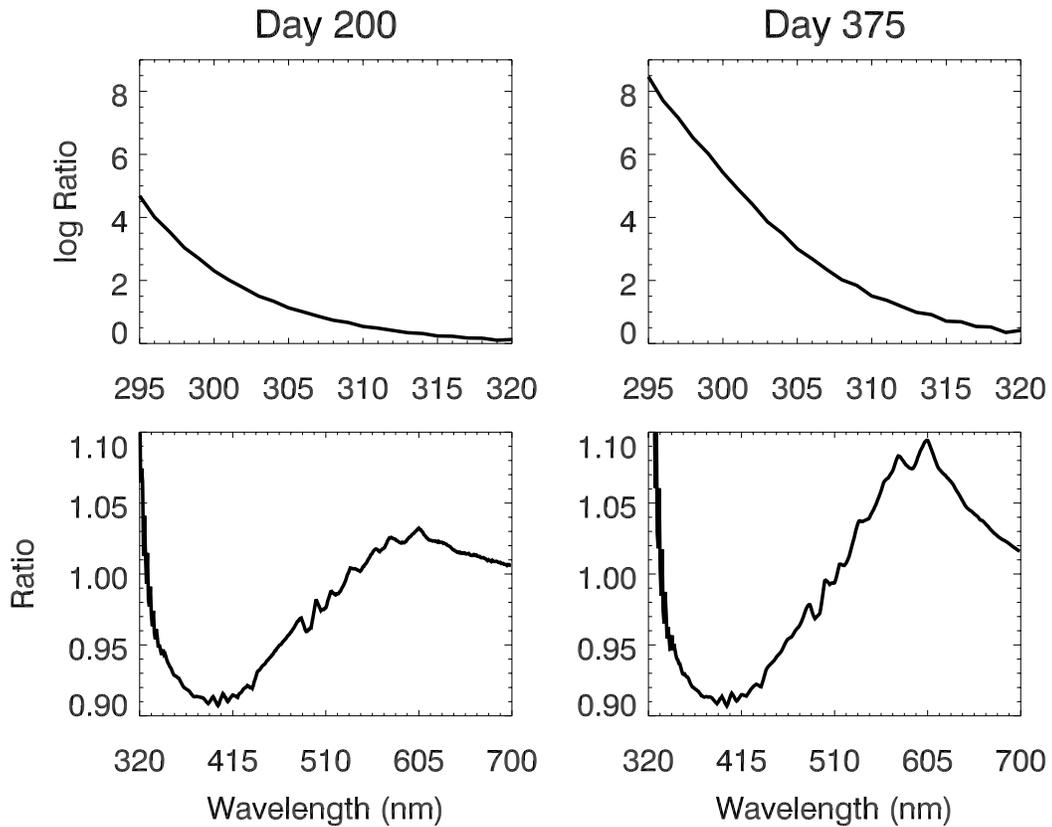

Figure 5

Ratio of spectral irradiance at Earth's surface in GRB case versus normal case, at 35°S latitude, on day 200 (left hand column) and day 375 (right hand column) following the GRB. Top panels show the log ratio in the UVB region of the spectrum; bottom panels show the simple ratio in the UVA and PAR regions of the spectrum.

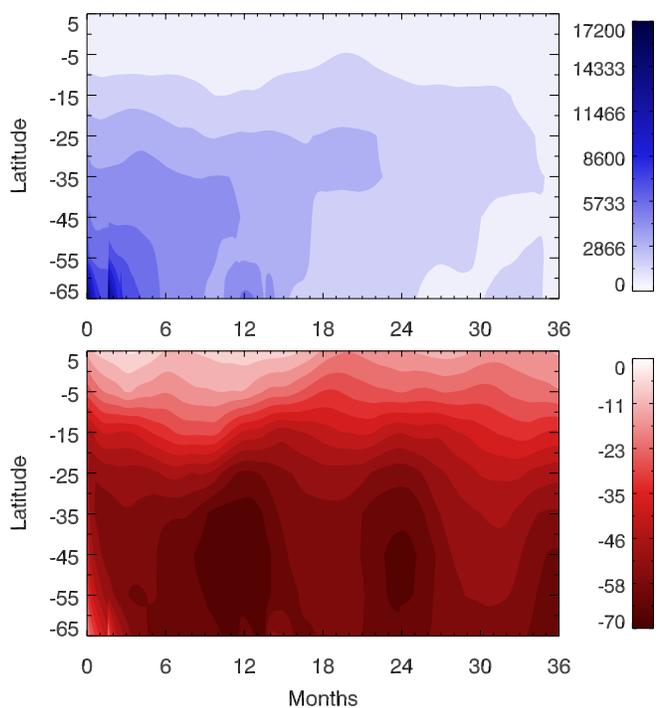

Figure 6

Pointwise percent difference in column density (comparing runs with and without GRB) of $NO_2$ (top panel) and $O_3$ (bottom panel).

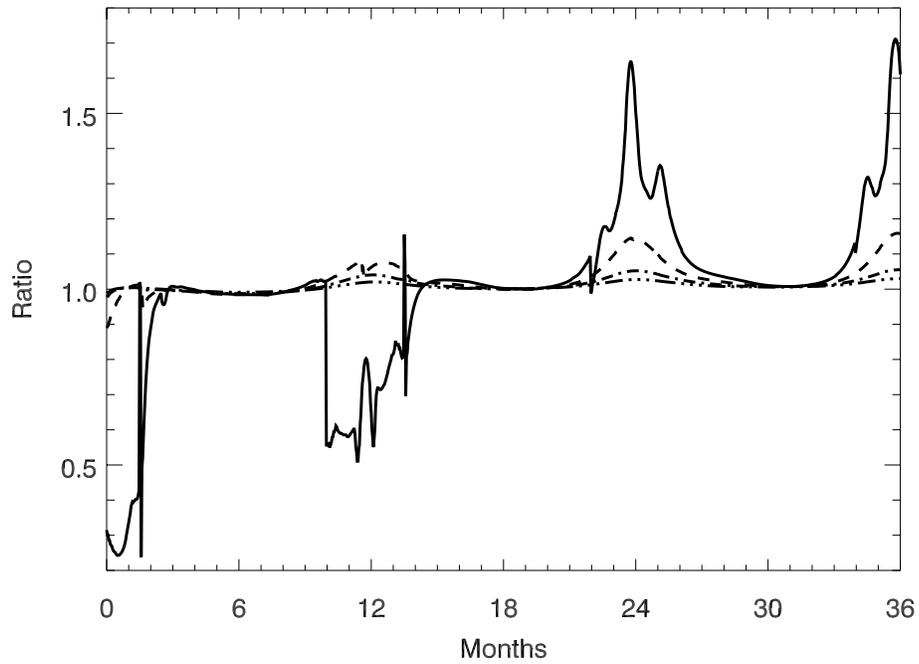

Figure 7

Ratio of PAR at Earth's surface in the GRB case versus the normal case, at latitudes 75°S (solid line), 65°S (dashed line), 55°S (dot-dash line), and 45°S (three-dot-dash line).

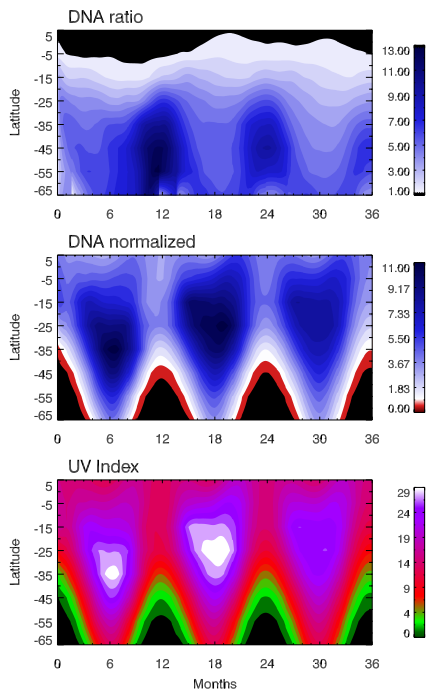

Figure 8

Top panel: Pointwise ratio of DNA damaging irradiance at Earth's surface in the GRB case versus that in the normal case. Middle panel: DNA damaging irradiance at Earth's surface in the GRB case, normalized by the annual global average damaging irradiance in the absence of a GRB. Bottom panel: UV Index at Earth's surface following a GRB.

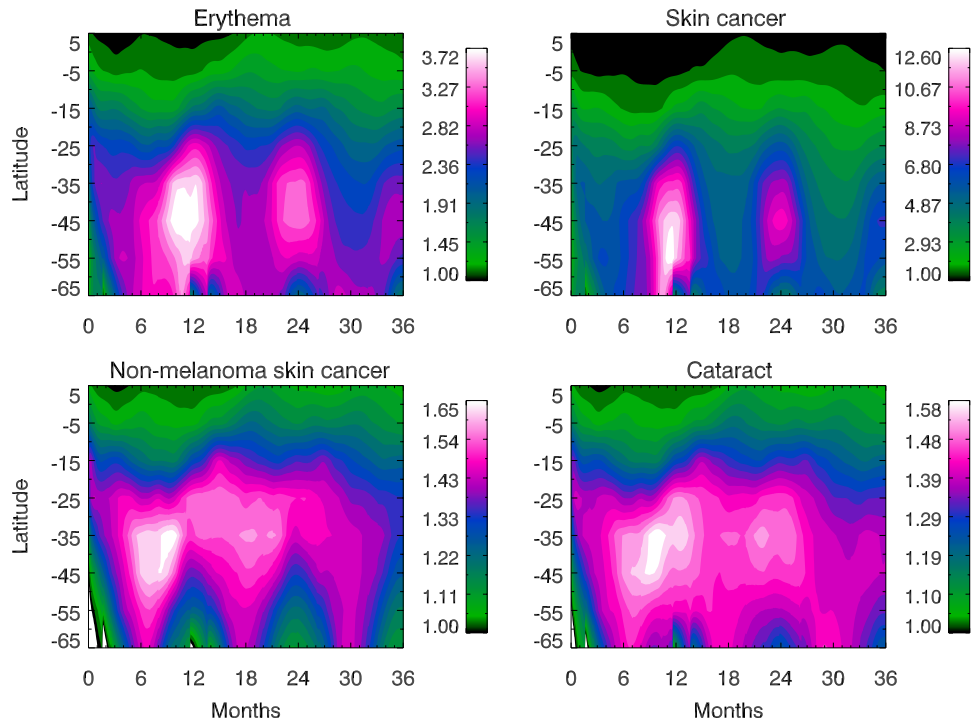

Figure 9

Pointwise ratios of computed biologically weighted irradiance at Earth's surface in the GRB case versus the normal case, using weighting functions for human erythema (top left panel); human skin cancer (top right panel); human non-melanoma skin cancer (bottom left panel); and cataracts in pigs (bottom right panel). Note that each panel has a different maximum value allowing better comparison of the time-space features of the increase in damaging irradiance.

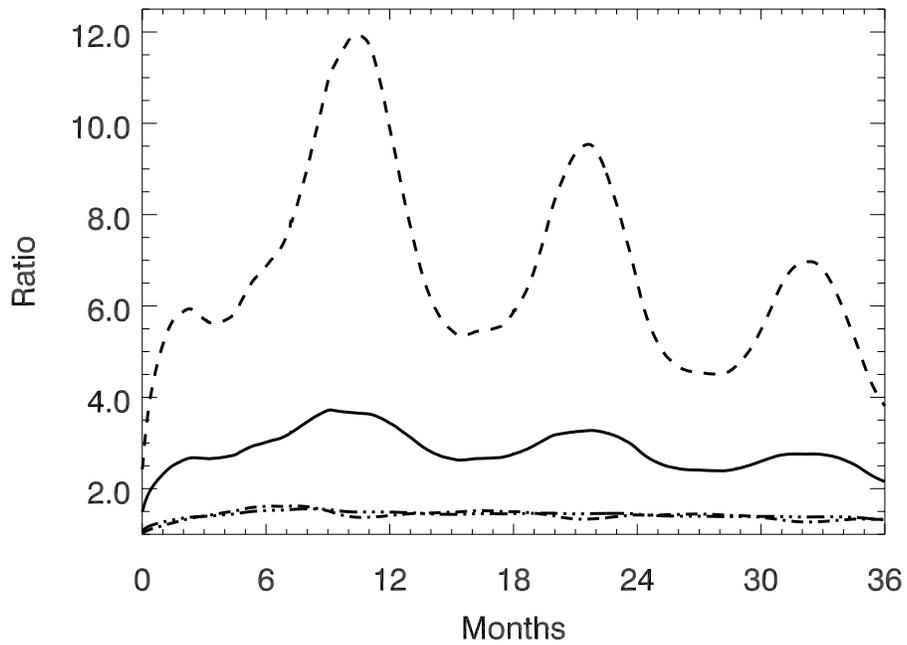

Figure 10

Ratios of computed biologically weighted irradiance at Earth's surface in the GRB case versus the normal case, at 45°S latitude for human erythema (solid line); human skin cancer (dashed line); human non-melanoma skin cancer (dot-dash line); and cataracts in pigs (three-dot-dash line).

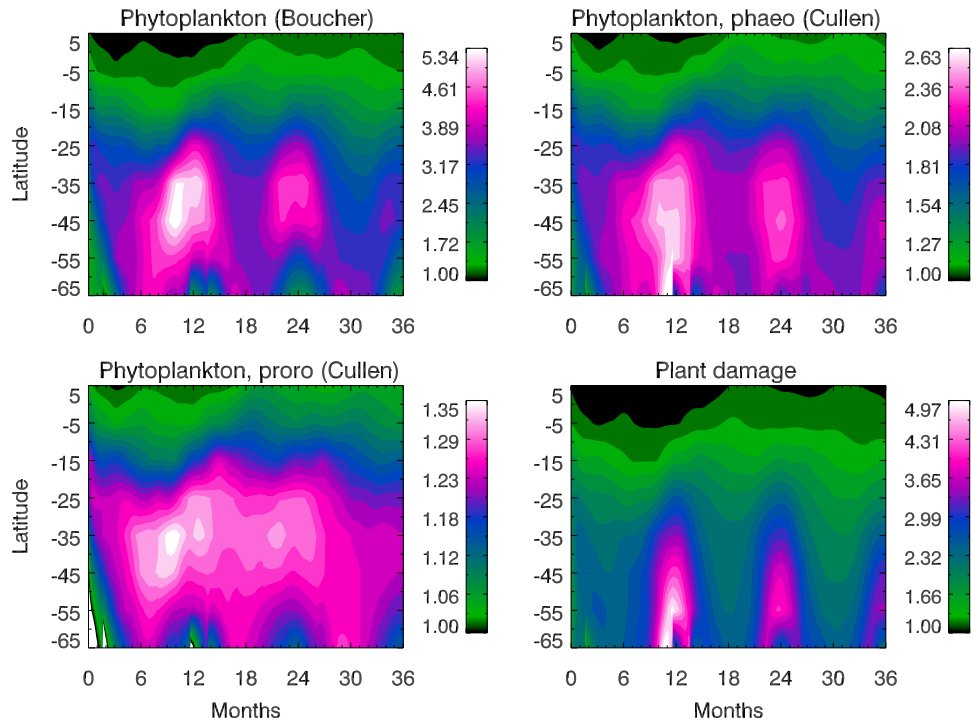

Figure 11

Pointwise ratios of biologically weighted irradiance at Earth's surface in the GRB case versus the normal case, computed using BWFs for inhibition of carbon fixation in Antarctic phytoplankton (top left panel); inhibition of photosynthesis in the diatom *Phaeodactylum* sp. (top right panel); inhibition of photosynthesis in the dinoflagellate *Prorocentrum micans* (bottom left panel); and inhibition of growth in oat seedlings (bottom right panel). Note that each panel has a different maximum value allowing better comparison of the time-space features of the increase in weighted irradiance.

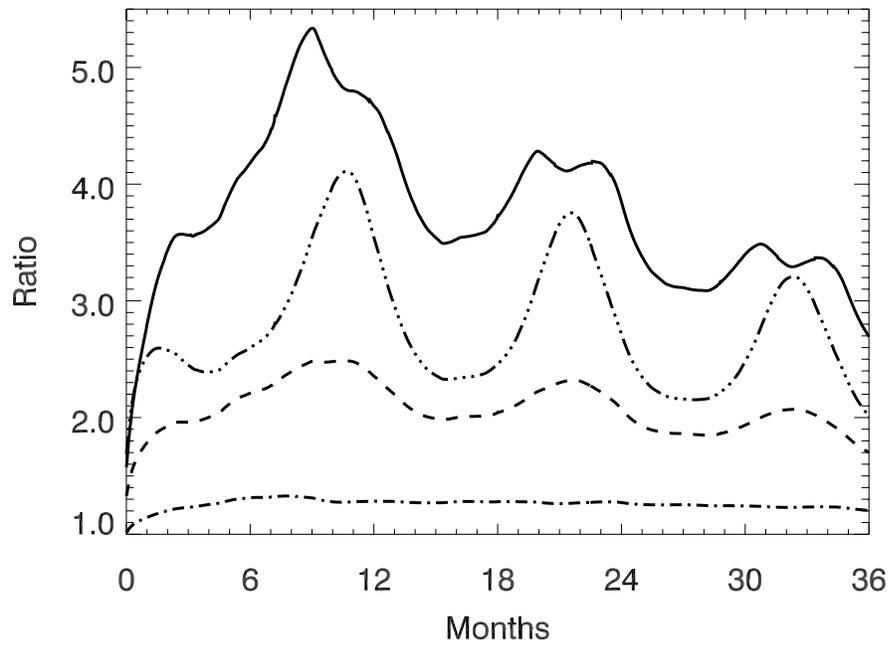

Figure 12

Ratios of computed biologically weighted irradiance at Earth's surface in the GRB case versus the normal case, at 45°S latitude using BWFs for inhibition of carbon fixation in Antarctic phytoplankton (solid line); inhibition of photosynthesis in the diatom *Phaeodactylum* sp. (dashed line); inhibition of photosynthesis in the dinoflagellate *Prorocentrum micans* (dot-dash line); and inhibition of growth in oat seedlings (three-dot-dash line).